\documentclass[12pt]{article}

\usepackage{amsmath,amssymb,amscd,amsbsy,amsgen,amsopn,amstext,
amsxtra}

\usepackage[dvips]{graphicx}

\begin{document}

\begin{flushright}
\end{flushright}

\vskip 0.5 truecm

\begin{center}
{\Large{\bf On the separability criterion for continuous 
variable systems}}
\end{center}
\vskip .5 truecm
\centerline{\bf  Kazuo Fujikawa }
\vskip .4 truecm
\centerline {\it Institute of Quantum Science, College of 
Science and Technology}
\centerline {\it Nihon University, Chiyoda-ku, Tokyo 101-8308, 
Japan}
\vskip 0.5 truecm

\makeatletter
\@addtoreset{equation}{section}
\def\theequation{\thesection.\arabic{equation}}
\makeatother

\begin{abstract}
We present an elementary and explicit proof of the 
separability criterion for continuous variable two-party 
Gaussian systems.
Our proof is based on an elementary formulation of uncertainty
relations and an explicit determination of squeezing parameters
for which  the P-representation 
condition  saturates the $Sp(2,R)\otimes Sp(2,R)$ invariant 
separability condition.
We thus give the explicit formulas of squeezing parameters, 
which establish the equivalence of the separability condition 
with the P-representation condition, in terms of the parameters of the standard form of the correlation matrix.   
Our proof is compared to the past proofs, and it is pointed out  
that the original proof of the P-representation by Duan, Giedke, Cirac and Zoller(DGCZ) is incomplete.
A way to complete their proof is then shown.
It is noted that both of the corrected 
proof of DGCZ and the proof of R. Simon are closely related to 
our explicit construction despite their quite different 
appearances. 
\end{abstract}

\section{Introduction}
The entanglement~\cite{epr} is an intriguing property of 
quantum mechanics, but a quantitative criterion of 
entanglement appears to be missing except for simple systems 
such as a two-spin system~\cite{peres, horodecki}. In view of 
this situation, it is 
remarkable that a quantitative sufficient condition for 
continuous variable two-party systems 
exists and that the criterion is necessary and sufficient for  
Gaussian states~\cite{duan, simon}. The proofs given by 
Duan, Giedke, Cirac and Zoller (DGCZ)~\cite{duan} and
Simon~\cite{simon} which consist of a series of logical steps
are ingenious. However, their proofs are based on some specific notions and ideas in quantum optics, and thus their proofs are not readily accessible to those physicists who are interested only in 
the general aspects of entanglement in quantum mechanics.
Moreover, these two proofs are seemingly quite different 
and their mutual connections are not
clear. This problem and related issues have been discussed in 
the past by 
several authors~\cite{englert, werner, giedke, mancini, eisert,
wolf, raymer, giovannetti}.
The present status of the quantum separability problem is nicely 
reviewed in~\cite{mancini2}.

We here present an elementary and explicit proof by starting 
with the elementary analysis of Heisenberg uncertainty 
relations in the manner of 
Kennard~\cite{kennard, robertson} and an explicit determination 
of squeezing parameters
which establish that the P-representation 
condition  saturates the $Sp(2,R)\otimes Sp(2,R)$ invariant 
separability condition. 
We thus give the explicit formulas of squeezing parameters, 
which establish the equivalence of the separability condition 
and the P-representation condition, in terms of the parameters of
the standard form of the correlation matrix (or second moments).
It is also pointed out that  
the original proof of the P-representation by DGCZ is incomplete,
and a way to complete their proof is 
shown. It is then shown that both of the corrected proof of DGCZ and the seemingly quite different proof of  Simon are closely related to our explicit construction.

Our treatment is based on a clear 
recognition that the separability condition 
associated with uncertainty relations is invariant under
general $Sp(2,R)\otimes Sp(2,R)$ transformations, whereas the 
condition for the P-representation of  Gaussian states is not 
invariant under general $Sp(2,R)\otimes Sp(2,R)$ 
transformations. A combination of these two apparently 
contradicting relations is the basis of our construction of the explicit solution.

\section{Entanglement and Kennard's relation}

\subsection{Kennard's relation}

We consider a two-party
system (or a two-particle system in one-dimensional space) 
described by canonical variables $(q_{1}, p_{1})$ and 
$(q_{2}, p_{2})$.
We define 
\begin{eqnarray}
&&\hat{X}(d,f)=d_{1}\hat{q}_{1}+d_{2}\hat{p}_{1}
+f_{1}\hat{q}_{2}+f_{2}\hat{p}_{2},\nonumber\\ 
&&\hat{X}(g,h)=g_{1}\hat{q}_{1}+g_{2}\hat{p}_{1}
+h_{1}\hat{q}_{2}+h_{2}\hat{p}_{2}
\end{eqnarray}
where all the coefficients
\begin{eqnarray}
d^{T}=(d_{1},d_{2}), \ \ f^{T}=(f_{1},f_{2}), 
\ \ g^{T}=(g_{1},g_{2}), \ \ h^{T}=(h_{1},h_{2})
\end{eqnarray}
are real numbers. The Kennard's relation for
a mixed state $\hat{\rho}=\sum_{k}P_{k}|\psi_{k}\rangle\langle
\psi_{k}|$ with $P_{k}\geq 0$ and $\sum_{k}P_{k}=1$ is
written as~\cite{simon} (for any choice of $d\sim h$) 
\begin{eqnarray}
\langle(\Delta\hat{X}(d,f))^{2}\rangle+
\langle(\Delta\hat{X}(g,h))^{2}\rangle\geq |d^{T}Jg+f^{T}Jh|
\end{eqnarray}
where we defined 
\begin{eqnarray}
\langle(\Delta\hat{X}(d,f))^{2}\rangle=
Tr\{(\Delta\hat{X}(d,f))^{2}\hat{\rho}\}
\end{eqnarray}
with
$\Delta\hat{X}(d,f)=\hat{X}(d,f)-\langle\hat{X}(d,f)\rangle$
and $\langle\hat{X}(d,f)\rangle=Tr\{\hat{X}(d,f)\hat{\rho}\}$,
for example, and the $2\times 2$ matrix
\begin{eqnarray}
J=\left(\begin{array}{cc}
  0&1\\
  -1&0\\
            \end{array}\right).
\end{eqnarray} 
The relation (2.3) is derived from 
$Tr\{\hat{\eta}\hat{\eta}^{\dagger}\hat{\rho}\}\geq0$ and 
$Tr\{\hat{\eta}^{\dagger}\hat{\eta}\hat{\rho}\}\geq0$ for
$\hat{\eta}=\Delta\hat{X}(d,f)+i\Delta\hat{X}(g,h)$, and
the right-hand side of (2.3) stands for $|[\Delta\hat{X}(d,f),
\Delta\hat{X}(g,h)]|$.

We examine (2.3) more precisely by starting with 
\begin{eqnarray}
\langle(\Delta\hat{X}(d,f))^{2}\rangle
&=&\langle\left(\hat{X}(d,f)-\langle\hat{X}(d.f)\rangle
\right)^{2}\rangle
\nonumber\\
&=&\sum_{k}P_{k}\langle\left(\hat{X}(d.f)
-\langle\hat{X}(d.f)\rangle\right)^{2}
\rangle_{k}\nonumber\\
&=&\sum_{k}P_{k}\langle\left(\hat{X}(d.f)
-\langle\hat{X}(d,f)\rangle_{k}
+\langle\hat{X}(d,f)\rangle_{k}
-\langle\hat{X}(d,f)\rangle\right)^{2}
\rangle_{k}\nonumber\\
&=&\sum_{k}P_{k}[\langle\left(\hat{X}(d.f)
-\langle\hat{X}(d,f)\rangle_{k}
\right)^{2}\rangle_{k}+\left(\langle\hat{X}(d,f)\rangle_{k}-
\langle\hat{X}(d,f)\rangle\right)^{2}]\nonumber\\
&\geq&\sum_{k}P_{k}\langle\left(\hat{X}(d,f)
-\langle\hat{X}(d,f)\rangle_{k}\right)^{2}
\rangle_{k}
\end{eqnarray}
which holds for a general mixed state for any real numbers
$d$ and $f$. The equality sign
holds only for 
\begin{eqnarray}
\langle\hat{X}(d,f)\rangle_{k}-
\langle\hat{X}(d,f)\rangle=0
\end{eqnarray}
 for all $k$ where $\langle\hat{X}(d,f)\rangle
=\sum_{k}P_{k}\langle\hat{X}(d,f)\rangle_{k}$ with
\begin{eqnarray} 
\langle\hat{X}(d,f)\rangle_{k}
=\int dq_{1}dq_{2}\psi^{\star}_{k}(q_{1},q_{2})\hat{X}(d,f)
\psi_{k}(q_{1},q_{2}).
\end{eqnarray}
The condition (2.7) is trivial for a pure state, but it imposes
a stringent condition on a mixed state. 
The Kennard's relation for a general pure state is 
given by (2.3) if one  sets $P_{k}=1$ for specific $k$ and 
others zero  
\begin{eqnarray}
&&\langle\left(\hat{X}(d,f)
-\langle\hat{X}(d,f)\rangle_{k}\right)^{2}
\rangle_{k}+\langle\left(\hat{X}(g,h)
-\langle\hat{X}(g,h)\rangle_{k}\right)^{2}
\rangle_{k}\nonumber\\
&&\geq|d^{T}Jg+f^{T}Jh|
\end{eqnarray}
for any $d\sim h$, which is also written as 
\begin{eqnarray}
&&t^{2}\langle\left(\hat{X}(d,f)
-\langle\hat{X}(d,f)\rangle_{k}\right)^{2}
\rangle_{k}+\langle\left(\hat{X}(g,h)
-\langle\hat{X}(g,h)\rangle_{k}\right)^{2}
\rangle_{k}\nonumber\\
&&-t|d^{T}Jg+f^{T}Jh|\geq 0
\end{eqnarray}
by replacing $d\rightarrow td, \ f\rightarrow tf$ for any real
$t$ and thus the discriminant gives the conventional form of
Kennard's relation.
The Kennard's relations for pure states imply
\begin{eqnarray}
&&\sum_{k}P_{k}[\langle\left(\hat{X}(d,f)
-\langle\hat{X}(d,f)\rangle_{k}\right)^{2}
\rangle_{k}+\langle\left(\hat{X}(g,h)
-\langle\hat{X}(g,h)\rangle_{k}\right)^{2}
\rangle_{k}]\nonumber\\
&&\geq|d^{T}Jg+f^{T}Jh|
\end{eqnarray}
for any $d\sim h$, which is more precise than (2.3) because of 
the removal of extra terms as in (2.6). 

\subsection{Separability condition}

For a separable pure state 
$\psi_{k}(q_{1},q_{2})=\phi_{k}(q_{1})\varphi_{k}(q_{2})$, 
we have
\begin{eqnarray}
\langle\left(\hat{X}(d,f)
-\langle\hat{X}(d,f)\rangle_{k}\right)^{2}
\rangle_{k}
&=&\langle\left(d_{1}\hat{q}_{1}
+d_{2}\hat{p}_{1}
-\langle d_{1}\hat{q}_{1}+d_{2}\hat{p}_{1}\rangle_{k}\right)
^{2}\rangle_{k}\nonumber\\
&&+\langle\left(f_{1}\hat{q}_{2}
+f_{2}\hat{p}_{2}
-\langle f_{1}\hat{q}_{2}+f_{2}\hat{p}_{2}
\rangle_{k}\right)^{2}\rangle_{k}
\end{eqnarray}
and similarly for $\langle\left(\hat{X}(g,h)
-\langle\hat{X}(g,h)\rangle_{k}\right)^{2}
\rangle_{k}$.
We  thus have
\begin{eqnarray}
&&[\langle\left(\hat{X}(d,f)
-\langle\hat{X}(d,f)\rangle_{k}\right)^{2}
\rangle_{k}+\langle\left(\hat{X}(g,h)
-\langle\hat{X}(g,h)\rangle_{k}\right)^{2}
\rangle_{k}]\nonumber\\
&&=[\langle\left(\hat{X}(d,0)
-\langle\hat{X}(d,0)\rangle_{k}\right)^{2}
\rangle_{k}+\langle\left(\hat{X}(g,0)
-\langle\hat{X}(g,0)\rangle_{k}\right)^{2}
\rangle_{k}\nonumber\\
&&\ \ \ \ \ \ \ \ \ +
\langle\left(\hat{X}(0,f)
-\langle\hat{X}(0,f)\rangle_{k}\right)^{2}
\rangle_{k}+\langle\left(\hat{X}(0,h)
-\langle\hat{X}(0,h)\rangle_{k}\right)^{2}
\rangle_{k}]\nonumber\\
&&\geq|d^{T}Jg| +
|f^{T}Jh|
\end{eqnarray}
which holds for any $d \sim h$.  Here we used (2.9)
for $f=h=0$ or $d=g=0$.
The equality sign holds only for 
\begin{eqnarray}
&&[\left(d_{1}\hat{q}_{1}
+d_{2}\hat{p}_{1}
-\langle d_{1}\hat{q}_{1}+d_{2}\hat{p}_{1}\rangle_{k}\right)
+i\left(g_{1}\hat{q}_{1}
+g_{2}\hat{p}_{1}
-\langle g_{1}\hat{q}_{1}+g_{2}\hat{p}_{1}
\rangle_{k}\right)]\phi_{k}(q_{1})=0,
\nonumber\\
&&[\left(f_{1}\hat{q}_{2}
+f_{2}\hat{p}_{2}
-\langle f_{1}\hat{q}_{2}+f_{2}\hat{p}_{2}\rangle_{k}\right)
+i\left(h_{1}\hat{q}_{2}
+h_{2}\hat{p}_{2}
-\langle h_{1}\hat{q}_{2}+ h_{2}\hat{p}_{2}
\rangle_{k}\right)]\varphi_{k}(q_{2})=0,
\nonumber\\
\end{eqnarray}
for suitable $d \sim h$ with $d^{T}Jg>0$ and $f^{T}Jh>0$.

We finally conclude from (2.6) and (2.13) for any separable 
density matrix
\begin{eqnarray}
&&\langle\left(\Delta\hat{X}(d,f)\right)^{2}\rangle+
\langle\left(\Delta\hat{X}(g,h)\right)^{2}\rangle
\nonumber\\
&&\geq
\sum_{k}P_{k}[\left(\langle\hat{X}(d,f)\rangle_{k}-
\langle\hat{X}(d,f)\rangle\right)^{2}
+\left(\langle\hat{X}(g,h)\rangle_{k}-
\langle\hat{X}(g,h)\rangle\right)^{2}]\nonumber\\
&&+|d^{T}Jg| +|f^{T}Jh|
\end{eqnarray}
for any $d \sim h$.
 
We next define the variables
$(\hat{\xi}_{\alpha})=(\hat{q}_{1}, \hat{p}_{1}, \hat{q}_{2}, 
\hat{p}_{2})$ and the $4\times 4$ correlation matrix $V$ by
\begin{eqnarray}
V=(V_{\alpha\beta}), \ \ \  
V_{\alpha\beta}=\frac{1}{2}
\langle\Delta\hat{\xi}_{\alpha}\Delta\hat{\xi}_{\beta}+
\Delta\hat{\xi}_{\beta}\Delta\hat{\xi}_{\alpha}\rangle
=\frac{1}{2}
\langle\{\Delta\hat{\xi}_{\alpha}, \Delta\hat{\xi}_{\beta}\}
\rangle
\end{eqnarray}
with $\Delta\hat{\xi}_{\alpha}=\hat{\xi}_{\alpha}
-\langle\hat{\xi}_{\alpha}\rangle$, 
which can be written in the form 
\begin{eqnarray}
V=\left(\begin{array}{cc}
  A&C\\
  C^{T}&B\\
            \end{array}\right)
\end{eqnarray}
where $A$ and $B$ are $2\times 2$ real symmetric matrices and 
$C$ is a $2\times 2$ real matrix. 
We also define 
\begin{eqnarray}
\tilde{V}=(\tilde{V}_{\alpha\beta}), \ \ \  
\tilde{V}_{\alpha\beta}=\sum_{k}P_{k}
\langle\Delta\hat{\xi}_{\alpha}\rangle_{k}
\langle\Delta\hat{\xi}_{\beta}\rangle_{k}
\end{eqnarray}
and
\begin{eqnarray}
\tilde{V}=\left(\begin{array}{cc}
  \tilde{A}&\tilde{C}\\
  \tilde{C}^{T}&\tilde{B}\\
            \end{array}\right)
\end{eqnarray}
where $\tilde{A}$ and $\tilde{B}$ are $2\times 2$ real symmetric
 matrices and 
$\tilde{C}$ is a $2\times 2$ real matrix. Both of $V$ and 
$\tilde{V}$ are non-negative. This quantity $\tilde{V}$ plays
a central role in the P-representation.

The basic relation (2.15) for separable states is then written as
\begin{eqnarray}
&&d^{T}Ad+f^{T}Bf+2d^{T}Cf+g^{T}Ag+h^{T}Bh+2g^{T}Ch\nonumber\\
&&\geq d^{T}\tilde{A}d+f^{T}\tilde{B}f+2d^{T}\tilde{C}f
+g^{T}\tilde{A}g+h^{T}\tilde{B}h+2g^{T}\tilde{C}h
\nonumber\\
&&+|d^{T}Jg| + |f^{T}Jh|.
\end{eqnarray}
while the  Kennard relation for general states is written as 
\begin{eqnarray}
&&d^{T}Ad+f^{T}Bf+2d^{T}Cf+g^{T}Ag+h^{T}Bh+2g^{T}Ch\nonumber\\
&&\geq d^{T}\tilde{A}d+f^{T}\tilde{B}f+2d^{T}\tilde{C}f
+g^{T}\tilde{A}g+h^{T}\tilde{B}h+2g^{T}\tilde{C}h
\nonumber\\
&&+|d^{T}Jg+f^{T}Jh|.
\end{eqnarray}
Note the difference between $|d^{T}Jg| + |f^{T}Jh|$ and 
$|d^{T}Jg+f^{T}Jh|$.

The antisymmetric commutator parts in 
\begin{eqnarray}
\langle\Delta\hat{\xi}_{\alpha}\Delta\hat{\xi}_{\beta}\rangle
=\frac{1}{2}
\langle\{\Delta\hat{\xi}_{\alpha}, \Delta\hat{\xi}_{\beta}\}
\rangle
+\frac{1}{2}
\langle[\Delta\hat{\xi}_{\alpha}, \Delta\hat{\xi}_{\beta}]\rangle
\end{eqnarray}
which may be added to $A$ and $B$ in (2.17) do not contribute 
to (2.20) since $d\sim h$ are all real.

Under the 
$S_{1}\otimes S_{2}\in Sp(2,R)\otimes Sp(2,R)$  transformations 
of
$(\hat{q}_{1},\hat{p}_{1})$ and $(\hat{q}_{2},\hat{p}_{2})$,
respectively, we have
\begin{eqnarray}
&&A\rightarrow S_{1}AS^{T}_{1},\ \ \ 
B\rightarrow S_{2}BS^{T}_{2},\ \ \ C\rightarrow S_{1}CS^{T}_{2}
\nonumber\\
&&\tilde{A}\rightarrow S_{1}\tilde{A}S^{T}_{1},\ \ \ 
\tilde{B}\rightarrow S_{2}\tilde{B}S^{T}_{2},\ \ \ \tilde{C}
\rightarrow S_{1}\tilde{C}S^{T}_{2}
\end{eqnarray}
which is equivalent to the transformation 
\begin{eqnarray}
d\rightarrow S^{T}_{1}d,\ \ \ 
f\rightarrow S^{T}_{2}f,\ \ \ g\rightarrow S^{T}_{1}g,
 \ \ \ h\rightarrow S^{T}_{2}h
\end{eqnarray}
in (2.20) if one recalls $J=S_{1}JS^{T}_{1},\ J=S_{2}JS^{T}_{2}$
; the inequality (2.20), which is valid for any $d\sim h$, holds
 after the transformation (2.24) and in this sense (2.20) is 
invariant under the above $Sp(2,R)\otimes Sp(2,R)$. To be 
precise, we do not use any property of the wave function under 
$Sp(2,R)\otimes Sp(2,R)$, and thus our $Sp(2,R)\otimes Sp(2,R)$ 
transformation is rather defined by (2.23) for given constant
matrices $A,\ B$ and $C$.

The difference between the two expressions in (2.20) and 
(2.21) appears when one 
replaces $B$ and $C$ by $S^{T}_{3}BS_{3}$ and $CS_{3}$ (and 
also $\tilde{B}$ and $\tilde{C}$ by 
$S^{T}_{3}\tilde{B}S_{3}$ and $\tilde{C}S_{3}$), respectively,  
with 
\begin{eqnarray}
S_{3}=\left(\begin{array}{cc}
  1&0\\
  0&-1\\
            \end{array}\right).
\end{eqnarray}
One can undo the replacements  in 
the first expression (2.20) by transformations 
$f\rightarrow S_{3}f$ and $h\rightarrow S_{3}h$, whereas it 
leads to $|d^{T}Jg-f^{T}Jh|$ in the second Kennard relation 
(2.21). The separability condition thus demands that the Kennard relation should hold both for the original system and for the system with the replacements of $B$ and $C$ by $S^{T}_{3}BS_{3}$ and $CS_{3}$, respectively, which may {\em a priori} be unphysical for inseparable systems. By using $S_{3}$, one can
adjust the signature of ${\rm det}C$ at one's will~\cite{simon}.

It is also useful to consider the separability condition (2.20)
with subsidiary conditions $g=J^{T}d$ and 
$h=J^{T}f$,
\begin{eqnarray}
&&d^{T}Ad+f^{T}Bf+2d^{T}Cf+d^{T}JAJ^{T}d+f^{T}JBJ^{T}f
+2d^{T}JCJ^{T}f\nonumber\\
&&\geq d^{T}\tilde{A}d+f^{T}\tilde{B}f
+2d^{T}\tilde{C}f+d^{T}J\tilde{A}J^{T}d
+f^{T}J\tilde{B}J^{T}f
+2d^{T}J\tilde{C}J^{T}f
\nonumber\\
&&+(d^{T}d+f^{T}f)
\end{eqnarray}
and with subsidiary conditions $g=J^{T}d$ and 
$h=-J^{T}f$ 
\begin{eqnarray}
&&d^{T}Ad+f^{T}Bf+2d^{T}Cf
+d^{T}JAJ^{T}d+f^{T}JBJ^{T}f
-2d^{T}JCJ^{T}f\nonumber\\
&&\geq d^{T}\tilde{A}d+f^{T}\tilde{B}f
+2d^{T}\tilde{C}f+d^{T}J\tilde{A}J^{T}d
+f^{T}J\tilde{B}J^{T}f
-2d^{T}J\tilde{C}J^{T}f
\nonumber\\
&&+(d^{T}d+f^{T}f).
\end{eqnarray}
For general inseparable  states in (2.21), we 
have the first condition (2.26) only if one wants to keep 
$(d^{T}d+f^{T}f)$ on the right-hand side in this form. The 
basic  
$Sp(2,R)\otimes Sp(2,R)$ invariance of uncertainty relations 
(2.20) and (2.21)
is lost in these conditions (2.26) and (2.27) with subsidiary conditions, 
but they have applications in the analysis of the 
P-representation.

\section{Separability and P-representation in Gaussian states}

\subsection{General analysis}

One can bring any given $V$ in (2.17) by $Sp(2,R)\otimes Sp(2,R)$
transformations to the standard form~\cite{duan, simon} (see
also Appendix A)
\begin{eqnarray}
V_{0}=\left(\begin{array}{cccc}
  a&0&c_{1}&0\\
  0&a&0&c_{2}\\
  c_{1}&0&b&0\\
  0&c_{2}&0&b\\            
  \end{array}\right).
\end{eqnarray}

One may understand the relations (2.20) and (2.21) ( and also
(2.26) and (2.27))
in two different ways:\\ 
(i) In the first interpretation, one may 
understand these relations (2.20) and (2.21) as an infinite set of uncertainty 
relations (and their variants) for any given constants 
$d\sim h$. In this 
interpretation, the relations (2.26) and (2.27) correspond
to the ones used by DGCZ~\cite{duan} if one chooses $d$ 
and $f$ suitably.\\
(ii) In the second interpretation of the relations (2.20) and 
(2.21), 
one may understand these relations holding for any choice of 
$d\sim h$ and thus imposing constraints on the allowed
ranges  of the elements $a, b, c_{1}, c_{2}$ of the standard 
form of $V_{0}$ in (3.1), for example. We adopt this second 
interpretation, which was also adopted by Simon~\cite{simon}. 
In this interpretation, the full relation (2.20)
is more restrictive than the  relations (2.26) and (2.27) with 
the {\em subsidiary conditions} on $d\sim h$. In 
other words, the elements $a, b, c_{1}, c_{2}$ which satisfy
(2.20) automatically satisfy (2.26) and (2.27), but not the 
other way around. In our analysis below, we interpret these 
relations as constraints on $c_{1}, c_{2}$ for fixed $a,b$.

The separability criterion is given by (2.20).
On the other hand, the P-representation depends on the 
condition (see Appendix B)
\begin{eqnarray}
V\geq \frac{1}{2}I
\end{eqnarray}
namely
\begin{eqnarray}
d^{T}Ad+f^{T}Bf+2d^{T}Cf\geq \frac{1}{2}(d^{T}d+f^{T}f)
\end{eqnarray}
for any $d\sim f$. 
By using a special property of the P-representation, namely,
a special property of the coherent state, 
one can also write (3.2) as 
\begin{eqnarray}
P^{-1}=\left(\begin{array}{cc}
  \tilde{A}&\tilde{C}\\
  \tilde{C}^{T}&\tilde{B}\\
            \end{array}\right)=V-\frac{1}{2}I\geq 0
\end{eqnarray}
or
\begin{eqnarray}
&&d^{T}\tilde{A}d+f^{T}\tilde{B}f
+2d^{T}\tilde{C}f\nonumber\\
&&=d^{T}Ad+f^{T}Bf+2d^{T}Cf
-\frac{1}{2}(d^{T}d+f^{T}f)\geq 0
\end{eqnarray}
for any $d\sim f$. Here $P^{-1}$ agrees with $\tilde{V}$
in (2.19). See Appendix B.

We first note that the P-representation implies the 
separability condition, since (3.5) also implies 
\begin{eqnarray}
&&g^{T}\tilde{A}g
+h^{T}\tilde{B}h
+ 2g^{T}\tilde{C}h
\nonumber\\
&&=g^{T}Ag+h^{T}Bh
+ 2g^{T}Ch -\frac{1}{2} (g^{T}g+h^{T}h)\geq 0
\end{eqnarray}
and thus adding these two relations (3.5) and (3.6), we have 
\begin{eqnarray}
&&d^{T}Ad+f^{T}Bf+2d^{T}Cf+g^{T}Ag+h^{T}Bh+2g^{T}Ch\nonumber\\
&&= d^{T}\tilde{A}d+f^{T}\tilde{B}f+2d^{T}\tilde{C}f
+g^{T}\tilde{A}g+h^{T}\tilde{B}h+2g^{T}\tilde{C}h
\nonumber\\
&&\frac{1}{2}(d^{T}d+f^{T}f)+\frac{1}{2} (g^{T}g+h^{T}h).
\end{eqnarray}
When one combines this relation with 
\begin{eqnarray}
\frac{1}{2}(d^{T}d+f^{T}f)+\frac{1}{2} (g^{T}g+h^{T}h)
&\geq& \sqrt{(d^{T}d+f^{T}f) (g^{T}g+h^{T}h)}\nonumber\\
&=& \sqrt{(d^{T}d+f^{T}f) (g^{T}J^{T}Jg+h^{T}J^{T}Jh)}
\nonumber\\
&\geq&
|d^{T}Jg|+|f^{T}Jh|,
\end{eqnarray}
one reproduces the separability condition (2.20). This is 
natural since the P-representation is in fact separable.

But the inverse is not obvious. The separability condition 
(2,20) is invariant under $Sp(2,R)\otimes Sp(2,R)$ in (2.23), 
whereas
the P-representation condition  (3.2) or (3.3) is not invariant under 
$Sp(2,R)\otimes Sp(2,R)$ by  noting that 
\begin{eqnarray}
d^{T}S_{1}^{T}S_{1}d+f^{T}S_{2}^{T}S_{2}f \neq 
d^{T}d+f^{T}f
\end{eqnarray}
in general. In this sense these two 
conditions cannot be equivalent to each other.
One may write the condition for P-representation as 
\begin{eqnarray}
V-\frac{1}{2}SS^{T}\geq 0
\end{eqnarray}
for a {\em suitable but arbitrary} $S\in Sp(2,R)\otimes Sp(2,R)$;
this relation implies the ordinary P-representation condition
$S^{-1}V(S^{-1})^{T}-\frac{1}{2}I\geq 0$ for a suitable
$Sp(2,R)\otimes Sp(2,R)$ transformed $S^{-1}V(S^{-1})^{T}$.
Written in the form (3.10), the condition for P-representation 
has a formally invariant meaning in the following sense.
For any  $S_{1}\in Sp(2,R)\otimes Sp(2,R)$, we have 
\begin{eqnarray}
S_{1}VS_{1}^{T}-\frac{1}{2}S_{1}SS^{T}S_{1}^{T}\geq 0
\end{eqnarray}
which is written as 
\begin{eqnarray}
V^{\prime}-\frac{1}{2}S^{\prime}(S^{\prime})^{T}\geq 0
\end{eqnarray}
with $V^{\prime}=S_{1}VS_{1}^{T}$ amd $S^{\prime}=S_{1}S
\in Sp(2,R)\otimes Sp(2,R)$.

\subsection{A new explicit proof}

We here present an explicit proof of the separability criterion
for continuous variable two-party Gaussian systems.
Our explicit construction gives the formulas of squeezing parameters, which establish the equivalence of the separability condition with the P-representation condition, in terms of the parameters of the standard form of the correlation matrix (3.1).

When one regards the separability condition as a constraint on the range of
$|c_{1}|$ and $|c_{2}|$ in the standard form $V_{0}$ (3.1),  
it is written as 
\begin{eqnarray}
&&4(ab-c_{1}^{2})(ab-c_{2}^{2})\geq (a^{2}+b^{2})+2|c_{1}c_{2}|
-\frac{1}{4},\nonumber\\
&&\sqrt{(2a-1)(2b-1)}\geq |c_{1}|+|c_{2}|
\end{eqnarray}
together with $a\geq 1/2$ and $b\geq 1/2$.
The conditions $a\geq 1/2$ and $b\geq 1/2$ are respectively 
derived by setting $f=h=0$ and $d=g=0$ in (2.20).
The first relation in (3.13), which was derived by 
Simon~\cite{simon}, corresponds to 
\begin{eqnarray}
4{\rm det}[V_{0}+\frac{i}{2}\left(\begin{array}{cc}
  J&0\\
  0&J\\            
  \end{array}\right)]\geq 0
\end{eqnarray}
up to a transformation $S_{3}$ in (2.25), and thus it is 
manifestly invariant under $Sp(2,R)\otimes Sp(2,R)$. The 
second condition in (3.13) is derived by the weaker conditions 
in (2.26) and (2.27) for the standard representation $V_{0}$, 
and it is used to exclude the nonsensical solutions of (3.13) 
with $c^{2}_{1}\rightarrow\infty$ and 
$c^{2}_{2}\rightarrow\infty$ for fixed $a$ and $b$.
The conditions (3.13) are equivalent to (2.20).

 The separability condition (3.13) is explicitly solved  as
\begin{eqnarray}
c_{1}^{2}
&\leq&\frac{1}{4t^{2}}\{[2ab(1+t^{2})+t]
-2\sqrt{a^{2}b^{2}(1-t^{2})^{2}+t(a+bt)(at+b)}\},\nonumber\\
c_{2}^{2}
&\leq&\frac{1}{4}\{[2ab(1+t^{2})+t]
-2\sqrt{a^{2}b^{2}(1-t^{2})^{2}+t(a+bt)(at+b)}\}\nonumber\\
\end{eqnarray}
for 
\begin{eqnarray}
0 \leq t\equiv |c_{2}|/|c_{1}|\leq 1
\end{eqnarray}
where we choose $|c_{2}|\leq |c_{1}|$ without loss of generality.
 
On the other hand, one may choose 
$S$ in (3.10) as 
\begin{eqnarray}
S(r_{1},r_{2})S^{T}(r_{1},r_{2})=\left(\begin{array}{cccc}
  1/r_{1}&0&0&0\\
  0&r_{1}&0&0\\
  0&0&1/r_{2}&0\\
  0&0&0&r_{2}\\            
  \end{array}\right)
\end{eqnarray}
with suitably chosen $r_{1}\geq 1$ and $r_{2}\geq 1$.
By choosing the standard form of $V_{0}$ in (3.1), the eigenvalues of 
$V_{0}-\frac{1}{2}S(r_{1},r_{2})S^{T}(r_{1},r_{2})$
are given by 
\begin{eqnarray}
&&(\lambda_{1})_{\pm}=\frac{1}{2}\{(a-\frac{1}{2r_{1}})+
(b-\frac{1}{2r_{2}})\pm
\sqrt{((a-\frac{1}{2r_{1}})-
(b-\frac{1}{2r_{2}}))^{2}+4c_{1}^{2}}\},\nonumber\\
&&(\lambda_{2})_{\pm}=\frac{1}{2}\{(a-\frac{1}{2}r_{1})+
(b-\frac{1}{2}r_{2})\pm
\sqrt{((a-\frac{1}{2}r_{1})-(b-\frac{1}{2}r_{2}))^{2}
+4c_{2}^{2}}\}.
\end{eqnarray}
The P-representation exists if $(\lambda_{1})_{\pm}\geq 0$
and $(\lambda_{2})_{\pm}\geq 0$, namely, if the following
two conditions are simultaneously satisfied 
\begin{eqnarray}
&&(a-\frac{1}{2r_{1}})(b-\frac{1}{2r_{2}})\geq 
c_{1}^{2},
\nonumber\\
&&
(a-\frac{1}{2}r_{1})(b-\frac{1}{2}r_{2})\geq c_{2}^{2}
\end{eqnarray}
together with $a\geq\frac{1}{2},\ b\geq \frac{1}{2}$, 
$(a-\frac{1}{2r_{1}})+(b-\frac{1}{2r_{2}})\geq 0$, and
$(a-\frac{1}{2}r_{1})+(b-\frac{1}{2}r_{2})\geq 0$.

When one regards (3.15) and (3.19) as constraints on the 
pair of variables 
\begin{eqnarray}
(c_{1}^{2}, c_{2}^{2})
\end{eqnarray}
for given $a$ and $b$,
the P-representation condition is more restrictive than 
the separability conditions, namely, the set of points 
$(c_{1}^{2}, c_{2}^{2})$ allowed by the P-representation condition (3.19) 
always satisfy the separability condition (3.15). To be precise,
we are working on the line defined by 
$t^{2}=c_{2}^{2}/c_{1}^{2}$.
We thus expect  that these two conditions can coincide only for 
the extremal value of the P-representation condition (3.19) with respect to 
$r_{1}$ and $r_{2}$ with fixed $t$. We show that this 
is indeed the case.
 
We thus want to prove 
\begin{eqnarray}
&&(a-\frac{1}{2r_{1}})(b-\frac{1}{2r_{2}(t,r_{1})})\nonumber\\
&=&\frac{1}{t^{2}}
[(a-\frac{1}{2}r_{1})(b-\frac{1}{2}r_{2}(t,r_{1}))]\\
&=&\frac{1}{4t^{2}}\{[2ab(1+t^{2})+t]
-2\sqrt{a^{2}b^{2}(1-t^{2})^{2}+t(a+bt)(at+b)}\}\nonumber
\end{eqnarray}
for a suitable $1\leq r_{1}\leq 2a$ (and $1\leq r_{2}\leq 2b$)
for any given $0\leq t\leq 1$ by regarding $r_{2}$ as a 
function of $r_{1}$ and $t$. By this way we establish that the 
separability condition (3.15) agrees with the P-representation condition (3.19) with a suitable $Sp(2,R)\otimes Sp(2,R)$ transformation.

We start with the equality in the left-hand side of (3.21)
\begin{eqnarray}
(a-\frac{1}{2r_{1}})(b-\frac{1}{2r_{2}(t,r_{1})})
&=&\frac{1}{t^{2}}(a-\frac{1}{2}r_{1})(b-\frac{1}{2}r_{2}(t,r_{1}))
\end{eqnarray}
and take the derivative of the both hand sides with respect to 
$r_{1}$ with fixed $t$. We then have
\begin{eqnarray}
&&
[\frac{1}{2r^{2}_{1}}(b-\frac{1}{2r_{2}(t,r_{1})})+
(a-\frac{1}{2r_{1}})\frac{1}{2r^{2}_{2}(t,r_{1})}
\frac{\partial r_{2}}{\partial r_{1}}]\nonumber\\
&=&\frac{1}{t^{2}}[-\frac{1}{2}(b-\frac{1}{2}r_{2}(t,r_{1}))-
(a-\frac{1}{2}r_{1})\frac{1}{2}\frac{\partial r_{2}}{\partial r_{1}}]
\end{eqnarray}
which is solved as 
\begin{eqnarray}
\frac{\partial r_{2}}{\partial r_{1}}
=-\frac{(b-\frac{1}{2}r_{2}(t,r_{1}))+
\frac{t^{2}}{r_{1}^{2}}(b-\frac{1}{2r_{2}(t,r_{1})})}
{(a-\frac{1}{2}r_{1})+\frac{t^{2}}{r^{2}_{2}(t,r_{1})}
(a-\frac{1}{2r_{1}})}.
\end{eqnarray}
We next consider the stationary point (or extremal) of 
\begin{eqnarray}
(a-\frac{1}{2}r_{1})(b-\frac{1}{2}r_{2}(t,r_{1}))
\end{eqnarray}
with fixed $t$, namely
\begin{eqnarray}
-\frac{1}{2}(b-\frac{1}{2}r_{2}(t,r_{1}))-
(a-\frac{1}{2}r_{1})\frac{1}{2}\frac{\partial r_{2}}
{\partial r_{1}}=0.
\end{eqnarray}
This relation combined with (3.24) gives rise to
\begin{eqnarray}
(b-\frac{1}{2}r_{2}(t,r_{1}))(a-\frac{1}{2r_{1}})
\frac{1}{r^{2}_{2}}=(a-\frac{1}{2}r_{1})
(b-\frac{1}{2r_{2}(t,r_{1})})\frac{1}{r^{2}_{1}}
\end{eqnarray}
The relations (3.22) and (3.27) give 
\begin{eqnarray}
&&r_{1}=\frac{\frac{r_{2}}{t}a+\frac{1}{2}}{a+\frac{1}{2}
\frac{r_{2}}{t}}, \ \ \
r_{2}=\frac{\frac{r_{1}}{t}b+\frac{1}{2}}{b+\frac{1}{2}
\frac{r_{1}}{t}}
\end{eqnarray}
which are symmetric in $r_{1}$ and $r_{2}$.
The relations in (3.28)  are solved as 
\begin{eqnarray}
&&r_{1}=\frac{1}{at+b}\{ab(1-t^{2})+
\sqrt{a^{2}b^{2}(1-t^{2})^{2}+t(a+bt)(at+b)}\},\nonumber\\
&&r_{2}=\frac{1}{a+bt}\{ab(1-t^{2})+
\sqrt{a^{2}b^{2}(1-t^{2})^{2}+t(a+bt)(at+b)}\}
\end{eqnarray}
with $0\leq t=|c_{2}|/|c_{1}|\leq 1$, which determine the
squeezing parameters.

We see from (3.29) that 
\begin{eqnarray}
r_{1}=r_{2}=1
\end{eqnarray}
for $t=1$, and 
\begin{eqnarray}
r_{1}=2a, \ \ \ \ r_{2}=2b
\end{eqnarray}
for $t=0$. One can also confirm
\begin{eqnarray}
&&\infty>\frac{r_{1}}{t}=\frac{a+bt}{-ab(1-t^{2})+
\sqrt{a^{2}b^{2}(1-t^{2})^{2}+t(a+bt)(at+b)}}\geq 1,\nonumber\\
&&\infty>\frac{r_{2}}{t}=\frac{at+b}{-ab(1-t^{2})+
\sqrt{a^{2}b^{2}(1-t^{2})^{2}+t(a+bt)(at+b)}}\geq 1
\end{eqnarray}
by noting $t(at+b)\leq (a+bt)$ and $t(a+bt)\leq (at+b)$ for 
$0\leq t \leq 1$ and the triangle inequality. By recalling 
(3.28), we thus conclude
\begin{eqnarray}
2a\geq r_{1}\geq 1, \ \ \ \ 2b\geq r_{2}\geq 1
\end{eqnarray}
for $a\geq \frac{1}{2}$ and $b\geq \frac{1}{2}$.

We finally evaluate by using $r_{1}$ and $r_{2}$ in (3.29)
\begin{eqnarray}
&&\frac{1}{t^{2}}(a-\frac{1}{2}r_{1})(b-\frac{1}{2}r_{2})
\nonumber\\
&&=\frac{1}{4t^{2}}[a+at(\frac{a+bt}{at+b})-
\frac{\sqrt{a^{2}b^{2}(1-t^{2})^{2}+t(a+bt)(at+b)}}{at+b}]
\nonumber\\
&&\times[b+bt(\frac{at+b}{a+bt})-
\frac{\sqrt{a^{2}b^{2}(1-t^{2})^{2}+t(a+bt)(at+b)}}{a+bt}]
\nonumber\\
&&=\frac{1}{4t^{2}}\{[2ab(1+t^{2})+t]
-2\sqrt{a^{2}b^{2}(1-t^{2})^{2}+t(a+bt)(at+b)}\}
\end{eqnarray}
which is a remarkable identity. This relation establishes 
(3.21), namely,
the fact that the boundaries of the conditions for separability 
and P-representation coincide for any 
$0\leq t=|c_{2}|/|c_{1}|\leq 1$.  

Our explicit construction proves that  the 
P-representation condition (3.19)  with suitably chosen 
$S(r_{1},r_{2})\in Sp(2,R)\otimes Sp(2,R)$, where
$1\leq r_{1}\leq 2a$ and $1\leq r_{2}\leq 2b$, is equivalent 
to the separability condition (3.15) for any 
$0\leq t=|c_{2}|/|c_{1}|\leq 1$, and thus the separability 
condition (3.15) is a necessary and sufficient separability
criterion for two-party Gaussian systems. Our formulas of 
 $r_{1}$ and $r_{2}$ in (3.29) give the explicit expressions of 
squeezing parameters to achieve the above equivalence in terms 
of the parameters of the standard form of the correlation matrix
 (3.1).

\section{Comparison with the past proofs}

\subsection{Proof of Duan,Giedke,Cirac and Zoller }

The analysis of Duan, Giedke, Cirac and Zoller 
(DGCZ)~\cite{duan} starts with the constraint
\begin{eqnarray}
\frac{(\frac{2a}{r_{1}}-1)}{(2ar_{1}-1)}=
\frac{(\frac{2b}{r_{2}}-1)}{(2br_{2}-1)}
\end{eqnarray}
which is written as 
\begin{eqnarray}
\frac{(\frac{n}{r_{1}}-1)}{(nr_{1}-1)}
=\frac{(\frac{m}{r_{2}}-1)}{(mr_{2}-1)}
\end{eqnarray}
by noting $2a=n$ and $2b=m$ in their notation of the 
$Sp(2,R)\otimes Sp(2,R)$ transformed correlation matrix
\begin{eqnarray}
M=\left(\begin{array}{cccc}
  nr_{1}&0&c\sqrt{r_{1}r_{2}}&0\\
  0&n/r_{1}&0&c^{\prime}/\sqrt{r_{1}r_{2}}\\
  c\sqrt{r_{1}r_{2}}&0&mr_{2}&0\\
  0&c^{\prime}/\sqrt{r_{1}r_{2}}&0&m/r_{2}\\            
  \end{array}\right)
\end{eqnarray}
with $n>m\geq 1$ and $|c|\geq |c^{\prime}|>0$. Their 
normalization corresponds to $M=2V$. 

One can rewrite (4.2) as 
$r_{2}(mr_{2}-1)=X(r_{1})(m-r_{2})$ and solve this quadratic 
equation in $r_{2}$ in the form
\begin{eqnarray}
r_{2}(r_{1})_{\pm}=\frac{1-X \pm\sqrt{(1-X)^{2}+4m^{2}X}}{2m}
\end{eqnarray}
where we defined 
\begin{eqnarray}
X(r_{1})=\frac{(nr_{1}-1)}{(\frac{n}{r_{1}}-1)}
=\frac{r_{1}(nr_{1}-1)}{n-r_{1}}
\end{eqnarray}
which assumes $X(1)=1$, $X(n-\epsilon)=\infty$, 
$X(n+\epsilon)=-\infty$ and $X(\infty)=-\infty$. Here 
$\epsilon$ is an infinitesimal positive quantity which is 
eventually set to $0$.
One thus finds
\begin{eqnarray}
&&r_{2}(1)_{+}=1, \ \ \ r_{2}(1)_{-}=-1,\nonumber\\
&&r_{2}(n-\epsilon)_{+}=m-\epsilon, \ \ \ r_{2}(n-\epsilon)_{-}
=-\infty,
\nonumber\\
&&r_{2}(n+\epsilon)_{+}=\infty, \ \ \ r_{2}(n+\epsilon)_{-}
=m+\epsilon,
\nonumber\\
&&r_{2}(\infty)_{+}=\infty, \ \ \ r_{2}(\infty)_{-}=m
\end{eqnarray}
and thus the solution has a rather involved branch structure
~\footnote{
By writing (4.5) as
\begin{eqnarray}
X(r_{1})\equiv
\frac{r_{1}(nr_{1}-1)}{n-r_{1}}
=-[n(r_{1}-n)+\frac{n(n^{2}-1)}{r_{1}-n}]-(2n^{2}-1)
\nonumber
\end{eqnarray}
one can show 
$-\infty< X(r_{1})\leq -[n+\sqrt{n^{2}-1}]^{2}$
for $n< r_{1}<\infty$ and the upper bound is achieved at
$r_{1}=n+\sqrt{n^{2}-1}$. Since $X>0$ for $1\leq 
r_{1}\leq n$, one can  confirm that the content
inside the square root in (4.4)
\begin{eqnarray}
(1-X)^{2}+4m^{2}X=\left(X+(m+\sqrt{m^{2}-1})^{2}\right)
\left(X+(m-\sqrt{m^{2}-1})^{2}\right)
\nonumber
\end{eqnarray}
is positive definite for $1\leq 
r_{1}<\infty$ when $n>m\geq 1$. Consequently the 
solutions $r_{2}(r_{1})_{\pm}$ in (4.4) are real, and 
$r_{2}(r_{1})_{+}$ for $1\leq r_{1}\leq n$ and 
$r_{2}(r_{1})_{-}$ for $n<r_{1}<\infty$ defines a continuous 
real function $r_{2}(r_{1})$ for $1\leq r_{1}<\infty$, which is 
all that is necessary in the analysis in~\cite{duan}. 
Incidentally, 
the content inside the square root above becomes negative and 
thus $r_{2}(r_{1})_{\pm}$ become complex for 
$m>n$ for some interval in  $n< r_{1}<\infty$. In particular 
 for $n=m$, $r_{2}=r_{1}$ is a solution.  

When one rewrites (4.4) as 
\begin{eqnarray}
r_{2}(r_{1})_{\pm}\equiv
\frac{2m}{1-\frac{1}{X}\pm\sqrt{(1-\frac{1}{X})^{2}
+4m^{2}\frac{1}{X}}}
\nonumber
\end{eqnarray}
for $X>0$ by taking $X$ inside the square root and continues 
this expression for negative $X$
also, then the single branch $r_{2}(r_{1})_{+}$ covers the 
entire domain $1\leq 
r_{1}<\infty$ for $n>m$. This is because one picks up an extra 
$-$ sign when one takes negative $X$ inside the square root. }.

One may next consider~\cite{duan} 
\begin{eqnarray}
f(r_{1})&=&\sqrt{r_{1}r_{2}}|c|-
\frac{|c^{\prime}|}{\sqrt{r_{1}r_{2}}}\nonumber\\
&&-[
\sqrt{(nr_{1}-1)(mr_{2}-1)}
-\sqrt{(\frac{n}{r_{1}}-1)(\frac{m}{r_{2}}-1)}\ ]
\end{eqnarray}
which satisfies (if one chooses the first branch 
$r_{2}(r_{1})_{+}$ in (4.4))
\begin{eqnarray}
f(1)=|c| - |c^{\prime}|\geq 0.
\end{eqnarray}
They then use the bound~\cite{duan}
\begin{eqnarray}
&&|c|\leq \sqrt{n(m-\frac{1}{m})}=\sqrt{n(m-1)\frac{m+1}{m}},
\end{eqnarray}
which is derived from (2.26) and (2.27) by setting $g=0$.
One may now examine $f(r_{1})$ in (4.7) in the domain 
$n< r_{1}<\infty$~\cite{duan} by choosing the branch 
$r_{2}(r_{1})_{-}$ in (4.4) for which one can show 
$m< r_{2}(r_{1})_{-}$. One then establishes
 $f(\infty)\leq 0$ by using (4.9) as in~\cite{duan}.

One thus concludes a solution for $f(r_{1})=0$ in the interval
$1\leq r_{1}<\infty$, as shown in~\cite{duan}, for the set of states which 
satisfy the condition (4.9).
Incidentally, 
\begin{eqnarray}
f(n)=\sqrt{nm}|c|-\frac{|c^{\prime}|}{\sqrt{nm}}-
\sqrt{(n^{2}-1)(m^{2}-1)},
\end{eqnarray}
and thus if one can establish
\begin{eqnarray}
\sqrt{nm}|c|-\frac{|c^{\prime}|}{\sqrt{nm}}\leq \sqrt{(n^{2}-1)
(m^{2}-1)},
\end{eqnarray}
one then has $f(n)\leq 0$. So far we briefly summarized the 
analysis in~\cite{duan} together with a comment on (4.10) and (4.11) which are used later.  

We now examine the main issue of the separability condition, 
eq.(16) in~\cite{duan},
\begin{eqnarray}
a_{0}^{2}\frac{n_{1}+n_{2}}{2}+\frac{m_{1}+m_{2}}{2a_{0}^{2}}
-|c_{1}|-|c_{2}|\geq a_{0}^{2}+\frac{1}{a_{0}^{2}}
\end{eqnarray}
with $a_{0}^{2}=\sqrt{\frac{m_{1}-1}{n_{1}-1}}
=\sqrt{\frac{m_{2}-1}{n_{2}-1}}$, which gives 
rise to
\begin{eqnarray}
\sqrt{(m_{1}-1)(n_{1}-1)}\pm \sqrt{(m_{2}-1)(n_{2}-1)}\geq
|c_{1}|+|c_{2}|
\end{eqnarray}
for $1\leq r_{1}\leq n$ and $n< r_{1}$, respectively, if one 
recalls that $n_{2}<1$ and $m_{2}<1$ for $n< r_{1}$
when (4.3) is written in the form
\begin{eqnarray}
M=\left(\begin{array}{cccc}
  n_{1}&0&c_{1}&0\\
  0&n_{2}&0&c_{2}\\
  c_{1}&0&m_{1}&0\\
  0&c_{2}&0&m_{2}\\            
  \end{array}\right).
\end{eqnarray}
Note the appearance of the crucial $\pm$ sign in (4.13) because of $(n_{2}-1)\sqrt{\frac{m_{2}-1}{n_{2}-1}}=-|n_{2}-1|
\sqrt{\frac{m_{2}-1}{n_{2}-1}}=-\sqrt{(m_{2}-1)(n_{2}-1)}$ for
$n_{2}<1$, for example. It appears that this minus sign was 
overlooked in~\cite{duan}.  

If  $f(r_{1})=0$ has a solution in the interval 
$1\leq r_{1}\leq n$, one can derive 
the condition for the P-representation (eq.(17) in~\cite{duan}) 
\begin{eqnarray}
\sqrt{(m_{1}-1)(n_{1}-1)}\geq |c_{1}|, \ \ \ 
\sqrt{(m_{2}-1)(n_{2}-1)}\geq |c_{2}|
\end{eqnarray}
by combining the first relation in (4.13) with  $f(r_{1})=0$,
and the proof of the P-representation in~\cite{duan}
naturally goes through. On the other hand, if
 $f(r_{1})=0$ has a solution in the interval 
$n< r_{1}<\infty$, one finds that the separability 
condition (4.13) with $n< r_{1}$  is 
inconsistent with $f(r_{1})=0$ for $|c_{2}|\neq 0$ since 
one then has
\begin{eqnarray}
\sqrt{(m_{1}-1)(n_{1}-1)}- \sqrt{(m_{2}-1)(n_{2}-1)}
=|c_{1}|-|c_{2}|\geq|c_{1}|+|c_{2}|.
\end{eqnarray}
This puzzling result for $n< r_{1}$ may indicate that 
some essential information is missing to analyze the 
P-representation. This is indeed the case as shown below. 

The condition $M-I\geq 0$  of the P-representation in fact 
requires 
\begin{eqnarray}
(n_{1}-1)+(m_{1}-1)\geq 0, \ \ \ (n_{2}-1)+(m_{2}-1)\geq 0
\end{eqnarray}
in addition to (4.16), since  the eigenvalues of $M-I$
are given by 
\begin{eqnarray}
(\lambda_{1})_{\pm}&=&\frac{1}{2}[(n_{1}-1)
+(m_{1}-1)\pm \sqrt{((n_{1}-1)
-(m_{1}-1))^{2}+4c^{2}_{1}}],\nonumber\\ 
(\lambda_{2})_{\pm}&=&\frac{1}{2}[(n_{2}-1)
+(m_{2}-1)\pm \sqrt{((n_{2}-1)
-(m_{2}-1))^{2}+4c^{2}_{2}}].
\end{eqnarray}
See (3.18) in a different representation.
 It is obvious 
that if (4.17) are not satisfied, at least one of the 
eigenvalues in (4.18) becomes negative. More intuitively, 
$M-I\geq 0$ cannot be maintained for sufficiently small $c_{1}$ 
and $c_{2}$ if (4.17) are not satisfied.
One can confirm that (4.15) is derived from the requirement  
$(\lambda_{1,2})_{\pm}\geq 0$ in (4.18), which is equivalent to 
$M-I\geq 0$, only under the conditions (4.17).
These extra conditions (4.17) are missing 
in~\cite{duan}, and the condition $(n_{2}-1)+(m_{2}-1)\geq 0$
is violated in the case with $n< r_{1}<\infty$ for 
which $n_{2}<1$ and $m_{2}<1$, and thus no P-representation   
exists for $n< r_{1}<\infty$~\footnote{It appears that for most 
cases in practice,
the continuous variable states automatically satisfy Lemma 2 
(standard form II) in~\cite{duan}, namely, the matrix $M$ in 
(4.3) with the constraint (4.2) for a suitable
$r_{1}$ in the interval $1\leq r_{1}<\infty$ which satisfies 
$f(r_{1})=0$ in (4.7). It should be 
emphasized that standard form II in~\cite{duan} as it stands is 
based on 
the quite weak condition (4.9) and thus it is valid for 
 inseparable states also. The standard form II holds even for 
$n< r_{1}$ but the P-representation does
not exist for $n< r_{1}$ and $m< r_{2}$, namely, for
$n_{2}<1$ and $m_{2}<1$.}. 

It has been argued in~\cite{duan} that separability or 
inseparability is independent of squeezing, but when one replaces
separability by the P-representation one finds that the
P-representation is very sensitive to squeezing.
No information about (4.17) is contained either in the 
separability condition (4.12)(eq.(16) in~\cite{duan}
in terms of EPR-like operators) or in their analysis of 
$f(r_{1})=0$, and in this sense one may conclude that the 
proof the P-representation in the original scheme of~\cite{duan} is 
incomplete. 

One may now recall that the exact separability condition (3.14) 
is $Sp(2,R)\otimes Sp(2,R)$ invariant while the P-representation 
condition is not invariant as is shown in (3.9). The 
separability condition
(3.15) is thus independent of squeezing parameters while
the P-representation condition (3.19) explicitly depends on 
squeezing parameters. The squeezing is an auxiliary device 
to show that the P-representation condition combined with  
suitable squeezing is equivalent to the 
$Sp(2,R)\otimes Sp(2,R)$ invariant separability condition. 
A salient feature of the analysis in~\cite{duan} is that they 
use the separability condition (4.12) which depends on squeezing
 parameters. 

To prove the P-representation 
starting with the separability condition, one needs to satisfy
two conditions (4.17) and (4.15). 
A way to achieve this purpose in the framework of ~\cite{duan}
is to show (4.15) by using (4.12) for the squeezing parameter in
 the range  
\begin{eqnarray}
1\leq r_{1}\leq n.
\end{eqnarray}
 We then automatically ensure (4.17) by 
means of (4.2) (and (4.6)) which is always assumed in our 
analysis. The non-trivial task is to show (4.15). For this 
purpose, we start with the first relation in (4.13), 
which is derived from the separability condition (4.12) if 
(4.2) is satisfied for $1\leq r_{1}\leq n$. 
It is then confirmed that the first relation in (4.13) 
with $r_{1}=n$ (and thus $r_{2}=m$ because of (4.2))
\begin{eqnarray}
\sqrt{(m^{2}-1)(n^{2}-1)}\geq
|c_{1}|+|c_{2}|=\sqrt{nm}|c|+\frac{|c^{\prime}|}{\sqrt{nm}}
\end{eqnarray}
ensures $f(n)\leq 0$ in (4.10). We thus conclude 
that there exists a solution for $f(r_{1})=0$ in the interval
 $1\leq r_{1}\leq n$ for the state which satisfies the 
separability condition (4.12) for any $r_{1}$ in 
 $1\leq r_{1}\leq n$. 
The proof of  the condition   
$M-I\geq 0$ for the P-representation is then complete.
Namely, (4.15) (eq.(17) in~\cite{duan}) together with
(4.17) is established by combining the first inequality in (4.13)
with $f(r_{1})=0$ for any $|c|\geq |c^{\prime}|$ in (4.3).
An important new ingredient of the present scheme compared to 
the original scheme in~\cite{duan} is that the order of the 
analyses of $f(r_{1})=0$ and the separability 
condition (4.12) is reversed and the separability 
condition (4.12) with $1\leq r_{1}\leq n$ now plays a central 
role in the analysis of $f(r_{1})=0$.
 One can confirm that (4.15) 
is equivalent to (3.19) when converted into our notation.   
\\

We here add further comments on the scheme of DGCZ in view of 
our explicit construction in Section 3. 

Firstly,
it is interesting that their condition (4.2), of which origin is 
not clearly stated in~\cite{duan}, agrees with our extremal 
condition (3.27). 

Secondly, it is shown that the weaker forms of the separability 
condition, (2.26) and (2.27), when applied to the representation
 (4.14) give rise to the condition 
\begin{eqnarray}
\sqrt{[(n_{1}+n_{2})-2][(m_{1}+m_{2})-2]}
\geq |c_{1}|+|c_{2}|.
\end{eqnarray}
To be explicit, (2.26) gives
\begin{eqnarray}
\left(\begin{array}{cccc}
  n_{1}+n_{2}&0&c_{1}+c_{2}&0\\
  0&n_{1}+n_{2}&0&c_{1}+c_{2}\\
  c_{1}+c_{2}&0&m_{1}+m_{2}&0\\
  0&c_{1}+c_{2}&0&m_{1}+m_{2}\\            
  \end{array}\right) -2I\geq 0\nonumber
\end{eqnarray}
by taking $M=2V$ into account, and (2.27) gives
\begin{eqnarray}
\left(\begin{array}{cccc}
  n_{1}+n_{2}&0&c_{1}-c_{2}&0\\
  0&n_{1}+n_{2}&0&-c_{1}+c_{2}\\
  c_{1}-c_{2}&0&m_{1}+m_{2}&0\\
  0&-c_{1}+c_{2}&0&m_{1}+m_{2}\\            
  \end{array}\right) -2I\geq 0.\nonumber
\end{eqnarray}
One can also show
\begin{eqnarray}
\sqrt{[(n_{1}+n_{2})-2][(m_{1}+m_{2})-2]}
\geq \sqrt{[n_{1}-1][m_{1}-1]}+ \sqrt{[n_{2}-1][m_{2}-1]}
\end{eqnarray}
where the equality holds only when the condition (4.2) is 
satisfied. This relation (4.22) is established by considering 
\begin{eqnarray}
f(x)=\sqrt{[n_{1}+x(n_{2}-n_{1}))-1]
[m_{1}+x(m_{2}-m_{1}))-1]}
\end{eqnarray}
with 
\begin{eqnarray}
f^{''}(x)&=&-\frac{1}{4}[(n_{1}-1)(m_{2}-1)
-(n_{2}-1)(m_{1}-1)]^{2}\nonumber\\
&\times&
[m_{1}+x(m_{2}-m_{1}))-1]^{-3/2}
[n_{1}+x(n_{2}-n_{1}))-1]^{-3/2}< 0\nonumber\\
\end{eqnarray}
except for (4.2), namely,
\begin{eqnarray}
\frac{(n_{2}-1)}{(n_{1}-1)}=\frac{(m_{2}-1)}{(m_{1}-1)}\nonumber
\end{eqnarray}
for which $f^{''}(x)=0$. By using the property of the 
convex function $2f(1/2)\geq f(1)+f(0)$
one can establish (4.22); the condition $1\leq r_{1}\leq n$ is 
sufficient to keep $f(x)$ real for $0\leq x\leq 1$.
Their separability condition (4.13), which is derived from 
(4.21) when the equality in (4.22) holds, thus corresponds 
to the weaker form of the separability condition.

This fact suggests that under the extremal condition (4.2), 
the weaker forms of the separability condition, (2.16) and 
(2.27), are sufficient to ensure the P-representation if 
supplemented by an additional constraint $1\leq r_{1}\leq n$.

\subsection{Proof of Simon}

The analysis of the case $c_{1}c_{2}\geq 0$ by 
Simon~\cite{simon} is quite elegant.
Starting with the standard form of $V_{0}$ in (3.1) and applying
 a set of $Sp(2,R)\otimes Sp(2,R)$ transformations, he arrives 
at the form of  $V$
\begin{eqnarray}
V=\left(\begin{array}{cccc}
  ay^{2}x^{2}&0&c_{1}y^{2}&0\\
  0&a/(y^{2}x^{2})&0&c_{2}/y^{2}\\
  c_{1}y^{2}&0&by^{2}/x^{2}&0\\
  0&c_{2}/y^{2}&0&bx^{2}/y^{2}\\            
  \end{array}\right)
\end{eqnarray}
which is also written as 
\begin{eqnarray}
V=\left(\begin{array}{cccc}
  ar_{1}&0&c_{1}\sqrt{r_{1}r_{2}}&0\\
  0&a/r_{1}&0&c_{2}/\sqrt{r_{1}r_{2}}\\
  c_{1}\sqrt{r_{1}r_{2}}&0&br_{2}&0\\
  0&c_{2}/\sqrt{r_{1}r_{2}}&0&b/r_{2}\\            
  \end{array}\right)
\end{eqnarray}
by defining
\begin{eqnarray}
r_{1}=(xy)^{2}, \ \ \ \ \ r_{2}=y^{2}/x^{2}.
\end{eqnarray}
He uses the crucial condition
\begin{eqnarray}
\frac{c_{1}}{ax^{2}-b/x^{2}}=
\frac{c_{2}}{a/x^{2}-bx^{2}}
\end{eqnarray}
which allows the  diagonalization of $V$ 
by a $Sp(4,R)$ transformation.  This
$Sp(4,R)$ preserves the Kennard relation, and thus one can use 
the Kennard relation to show the P-representation.

The condition (4.28) is written as 
\begin{eqnarray}
x^{4}=\frac{r_{1}}{r_{2}}=\frac{c_{1}a+c_{2}b}{c_{2}a+c_{1}b}
=\frac{a+(c_{2}/c_{1})b}{(c_{2}/c_{1})a+b}
\end{eqnarray}
which implies (for $a\geq b$)
\begin{eqnarray}
1 \leq \frac{r_{1}}{r_{2}}\leq \frac{a}{b}.
\end{eqnarray}
It is interesting that the condition (4.29) agrees with 
our explicit construction (3.29).

Simon~\cite{simon} shows that $V$ in (4.25) can be diagonalized 
by an 
$Sp(4,R)$ transformation as $V^{\prime}={\rm diag}(\kappa_{+},
\kappa^{\prime}_{+}, \kappa_{-},\kappa^{\prime}_{-})$
with
\begin{eqnarray}
\kappa_{\pm}&=&\frac{1}{2}y^{2}
\{ax^{2}+b/x^{2}\pm\sqrt{(ax^{2}-b/x^{2})^{2}+4c_{1}^{2}}\}
,\nonumber\\
\kappa^{\prime}_{\pm}&=&\frac{1}{2}y^{-2}
\{a/x^{2}+bx^{2}
\pm\sqrt{(a/x^{2}-bx^{2})^{2}+4c_{2}^{2}}\}.  
\end{eqnarray}
The equality of two smaller eigenvalues 
$\kappa_{-}=\kappa^{\prime}_{-}$ is 
ensured if 
\begin{eqnarray}
y^{4}=r_{1}r_{2}&
=&\frac{a/x^{2}+bx^{2}-\sqrt{(a/x^{2}-bx^{2})^{2}+4c_{2}^{2}}}
{ax^{2}+b/x^{2}-\sqrt{(ax^{2}-b/x^{2})^{2}+4c_{1}^{2}}}
\nonumber\\
&=&\frac{a+b(r_{1}/r_{2})-\sqrt{(a-b(r_{1}/r_{2}))^{2}
+4c_{2}^{2}(r_{1}/r_{2})}}
{a(r_{1}/r_{2})+b-\sqrt{(a(r_{1}/r_{2})-b)^{2}+
4c_{1}^{2}(r_{1}/r_{2})}}
\end{eqnarray}
and the Kennard's relation (uncertainty relation), 
$\kappa_{-}\kappa^{\prime}_{-}\geq 1/4$, implies 
$\kappa_{-}=\kappa^{\prime}_{-}\geq 1/2$ which
in turn gives rise to
\begin{eqnarray}
&&(ar_{1}-\frac{1}{2})(br_{2}-\frac{1}{2})\geq 
c_{1}^{2}r_{1}r_{2},
\nonumber\\
&&(a/r_{1}-\frac{1}{2})(b/r_{2}-\frac{1}{2})\geq c_{2}^{2}/
(r_{1}r_{2})
\end{eqnarray}
or equivalently 
\begin{eqnarray}
&&(a-\frac{1}{2r_{1}})(b-\frac{1}{2r_{2}})\geq c_{1}^{2},
\nonumber\\
&&(a-\frac{1}{2}r_{1})(b-\frac{1}{2}r_{2})\geq c_{2}^{2}.
\end{eqnarray}
This last relation naturally agrees with the condition of 
the P-representation (3.19). The separability condition, which 
agrees 
with the Kennard's relation for $c_{1}c_{2}>0$, thus ensures the 
P-representation. 
The case $c_{1}c_{2}<0$ is equally treated by replacing 
$c_{1}$ and $c_{2}$ by $|c_{1}|$ and $|c_{2}|$, 
respectively~\cite{simon}.

Although we have not succeeded in proving (4.32) by using our 
explicit solution (3.29) due to technical complications, one can 
confirm that (4.29) combined with one of the relations (3.28) 
gives rise to our explicit solution (3.29). We thus believe 
that the solution given by Simon agrees with our explicit
construction.  

\section{Conclusion}

We have presented an elementary and explicit analysis of the 
separability criterion of continuous variable two-party
Gaussian systems. In particular, we derived the explicit formulas of 
squeezing parameters, which establish the equivalence of the 
separability condition with the P-representation condition, in terms of 
the parameters of the standard form of the correlation matrix 
(or second moments).
 In the course of our analysis, we corrected 
the shortcomings of the past proof of DGCZ~\cite{duan}.
Our explicit construction also clarified the 
basic equivalence of the past seemingly quite different 
proofs of the separability criterion~\cite{duan, simon} in the 
sense that  both 
of the past proofs are closely related to the present explicitly
constructed solution.\\

I thank K. Shiokawa for an informative discussion.    

\appendix

\section{Standard form of $V$}

We recall the elements 
of $Sp(2,R)$ 
\begin{eqnarray}
S=\left(\begin{array}{cc}
  \cos\theta&\sin\theta\\
  -\sin\theta&\cos\theta\\
            \end{array}\right), \ \ \ \ 
S=\left(\begin{array}{cc}
  x&0\\
  0&\frac{1}{x}\\
            \end{array}\right)
\end{eqnarray}
which satisfy $SJS^{T}=J$.  One can bring $V$ in (2.17) to the 
standard form 
\begin{eqnarray}
V=\left(\begin{array}{cccc}
  a&0&c_{1}&0\\
  0&a&0&c_{2}\\
  c_{1}&0&b&0\\
  0&c_{2}&0&b\\            
  \end{array}\right)
\end{eqnarray}
by suitable $Sp(2,R)\otimes Sp(2,R)$ 
transformations~\cite{duan,
simon}; real
symmetric $A$ and $B$ can be made diagonal by two-dimensional 
rotations with suitable parameters $\theta$ in (A.1) and 
then applying the second elements in (A.1) with suitable 
parameters $x$, $A$ and $B$ are 
made proportional to the unit matrix. After these 
transformations $C$ remains real. By applying a suitable 
two-dimensional orthogonal
transformation $S_{1}\otimes S_{2}$, which is an element of 
$Sp(2,R)\otimes Sp(2,R)$, we can diagonalize $C$ 
\begin{eqnarray}
S_{1}CS^{T}_{2}=\left(\begin{array}{cc}
  c_{1}&0\\
  0&c_{2}\\            
  \end{array}\right).
\end{eqnarray}
By this way we arrive at (A.2).

\section{P-representation}
 
We define the generating function of all the correlations
(or moments) of  dynamical variables by
\begin{eqnarray}
\chi(\lambda, \eta)={\rm Tr}(\hat{\rho}\exp\{
i(\lambda_{1}\hat{q}_{1}+
\lambda_{2}\hat{p}_{1}+
\eta_{1}\hat{q}_{2}+\eta_{2}\hat{p}_{2})\})
\end{eqnarray}
where $\lambda_{1}\sim \eta_{2}$ are real numbers. By expanding 
$\chi(\lambda, \eta)$ in powers of $\lambda_{1}\sim \eta_{2}$, one can generate all the moments of dynamical variables.
Following the convention in this field, we define the Gaussian
states by 
\begin{eqnarray}
\chi(\lambda, \eta)=
\exp\{-\frac{1}{2}(\lambda_{1},\lambda_{2},\eta_{1},\eta_{2})V(\lambda_{1},\lambda_{2},\eta_{1},\eta_{2})^{T}\}
\end{eqnarray}
where $V$ is the correlation matrix in (2.17), namely, all the 
correlation functions are determined by the second moments.
One can write (B.1) as 
\begin{eqnarray}
\chi(\lambda, \eta)={\rm Tr}(\hat{\rho}\exp\{i(
\lambda^{\star}\hat{a}+\lambda\hat{a}^{\dagger}
+\eta^{\star}\hat{b}+\eta\hat{b}^{\dagger})\})
\end{eqnarray}
with
\begin{eqnarray}
&&\hat{a}=\frac{1}{\sqrt{2}}(\hat{q}_{1}+i\hat{p}_{1}),\ \ 
\hat{b}=\frac{1}{\sqrt{2}}(\hat{q}_{2}+i\hat{p}_{2}),\nonumber\\
&&\lambda=\frac{1}{\sqrt{2}}(\lambda_{1}+i\lambda_{2}),\ \ 
\eta=\frac{1}{\sqrt{2}}(\eta_{1}+i\eta_{2})
\end{eqnarray}
The Gaussian state is called P-representable if the density matrix is written as
\begin{eqnarray}
\hat{\rho}&=&\int d^{2}\alpha\int d^{2}\beta P(\alpha,\beta)
|\alpha,\beta\rangle\langle\alpha,\beta|
\end{eqnarray}
where $|\alpha,\beta\rangle$ is the coherent state defined by
\begin{eqnarray}
\hat{a}|\alpha,\beta\rangle=\alpha|\alpha,\beta\rangle,\ \ 
\hat{b}|\alpha,\beta\rangle=\beta|\alpha,\beta\rangle,\ \
\langle\alpha,\beta|\alpha,\beta\rangle=1
\end{eqnarray}
or to be explicit
\begin{eqnarray}
|\alpha,\beta\rangle=e^{\alpha\hat{a}^{\dagger}
-\frac{1}{2}|\alpha|^{2}}|0\rangle\otimes 
e^{\beta\hat{b}^{\dagger}-\frac{1}{2}|\beta|^{2}}|0\rangle.
\end{eqnarray}
Thus the P-representable states are separable.

By using the density matrix (B.5) in (B.3) and after normal 
ordering the exponential factor in (B.3), we have
\begin{eqnarray}
\chi(\lambda, \eta)&=&\int d^{2}\alpha\int d^{2}\beta P(\alpha,\beta)\exp\{i(
\lambda^{\star}\alpha+\lambda\alpha^{\star}
+\eta^{\star}\beta+\eta\beta^{\star})\}\nonumber\\
&&\times \exp\{-\frac{1}{2}(
|\lambda|^{2}+|\eta|^{2})\}
\end{eqnarray}
or, if one combines this expression with (B.2) we have
\begin{eqnarray}
&&\exp\{-\frac{1}{2}(\lambda_{1},\lambda_{2},\eta_{1},\eta_{2})
(V-\frac{1}{2}I)(\lambda_{1},\lambda_{2},\eta_{1},\eta_{2})^{T}\}
\nonumber\\
&&=\int d^{2}\alpha\int d^{2}\beta P(\alpha,\beta)\exp\{i(
\lambda_{1}\alpha_{1}+\lambda_{2}\alpha_{2}
+\eta_{1}\beta_{1}+\eta_{2}\beta_{2})\}
\end{eqnarray}
with $\alpha=(\alpha_{1}+i\alpha_{2})/\sqrt{2}$ and 
$\beta=(\beta_{1}+i\beta_{2})/\sqrt{2}$. Thus 
$P(\alpha,\beta)$ in (B.9) is given by
\begin{eqnarray}
P(\alpha,\beta)=\frac{\sqrt{det P}}{4\pi^{2}}
\exp\{-\frac{1}{2}(\alpha_{1},\alpha_{2},\beta_{1},\beta_{2})P(\alpha_{1},\alpha_{2},\beta_{1},\beta_{2})^{T}\}
\end{eqnarray}
where
\begin{eqnarray}
P^{-1}=V-\frac{1}{2}I\geq 0
\end{eqnarray}
which defines the condition for the P-representation.

The formula (B.9) indicates that the right-hand side 
generates the correlations of the form
\begin{eqnarray}
\int d^{2}\alpha\int d^{2}\beta P(\alpha,\beta)
\langle\alpha,\beta|\hat{a}^{\dagger}
|\alpha,\beta\rangle \langle\alpha,\beta|\hat{a}
|\alpha,\beta\rangle,
\end{eqnarray}
for example, which may be compared to (2.18). By recalling 
(B.5), this relation shows that all
the second moments in the right-hand side of (B.9)  
are given by $\tilde{V}$ in (2.18) (if one chooses 
$\langle \hat{a}^{\dagger}\rangle=\langle\hat{a}\rangle=
\langle \hat{b}^{\dagger}\rangle=\langle\hat{b}\rangle=0$).
This property establishes the special relation (3.4) of the 
P-representation.

\end{document}